\documentclass{appolb}
\usepackage{graphicx}
\usepackage{xspace}
\usepackage{bm}

\newcommand{\urs}{URu$_2$Si$_2$\xspace}

\newcommand{\moment}{$\mu_{\rm o}$\xspace}
\newcommand{\mb}{$\mu_{\rm B}/{\rm U}$\xspace}

\newcommand{\sigmatoa}{$\sigma\,||\,[100]$\xspace}

\newcommand{\sigmatoc}{$\sigma\,||\,[001]$\xspace}

\begin{document}
\title{Nonequilibrium Antiferromagnetic State in the Heavy Electron Compound URu$_2$Si$_2$
}


\author{M.\ Yokoyama$^{\rm a}$, J. Nozaki$^{\rm a}$, H.\ Amitsuka$^{\rm a}$, K.\ Watanabe$^{\rm b}$,\\ S.\ Kawarazaki$^{\rm b}$, H.\ Yoshizawa$^{\rm c}$ and J.A.\ Mydosh$^{\rm d}$
\address{
$^{\rm a}$Graduate School of Science, Hokkaido University, Sapporo 060-0810, Japan\\
$^{\rm b}$Graduate School of Science, Osaka University, Toyonaka 560-0043, Japan \\
$^{\rm c}$Neutron Scattering Laboratory, Institute for Solid State Physics,\\ University of Tokyo, Tokai 319-1106, Japan \\
$^{\rm d}$Kamerlingh Onnes Laboratory, Leiden University, P.O.Box 9504,\\ 2300 RA Leiden, The Netherlands
}
}
\maketitle


\begin{abstract}
We have investigated the nature of the antiferromagnetic (AF) phase induced by uniaxial stress $\sigma$ in \urs , by performing elastic neutron scattering measurements up to 0.4 GPa. We have found that the AF Bragg-peak intensity shows a clear hysteresis loop with $\sigma$ under the zero-stress cooling condition.
The result strongly suggests that the $\sigma$-induced AF phase is metastable and separated from the coexisting ``hidden ordered" phase by a first-order phase transition. We also present the analyses of the crystalline strain effects, and suggest that the $c/a$ ratio plays an important role in the competition between these two phases.  
\end{abstract}


The recent experimental findings of the unusual development of the antiferromagnetic (AF) phase~\cite{rf:Ami99,rf:Matsuda2001} induced by hydrostatic pressure have shed a new light on the issue of the ``hidden order" below $T_{\rm o}=17.5\ {\rm K}$ in \urs (the tetragonal ThCr$_2$Si$_2$-type structure~\cite{rf:Palstra85}). Neutron scattering measurements under hydrostatic pressure $P$ showed that the AF Bragg-peak intensities are strongly enhanced with increasing $P$~\cite{rf:Ami99}. Afterwards, $^{29}$Si-NMR experiments revealed that the volume of the AF state increases inhomogeneously by applying $P$, while that the magnitude of the ordered moment does not change~\cite{rf:Matsuda2001}. This indicates that the enhancement of the AF Bragg-peak intensities can be attributed to the increase of the AF volume fraction, not of the local AF moment. We have further investigated the effects of uniaxial stress $\sigma$ on these phases, and found that the AF Bragg-peak intensities show a significant increase for $\sigma$ applied along the tetragonal basal plane, while nearly $\sigma$-independent behavior for \sigmatoc ~\cite{rf:Yoko2002}. This is also considered to be caused by the inhomogeneous development of the AF phase, because the AF-moment value enhanced under $\sigma$, which is estimated by assuming the homogeneous AF order, is nearly equal to that observed under $P$. These experimental results strongly suggest that the two order parameters are separated into two independent phases by a first order phase transition. If so, hysteretic behavior is expected to be observed in $P$, $\sigma$ and $T$ variations of the AF state. However, this point is not clear in the previous measurements, where samples were cooled after the compression at room temperature. In this paper, we have performed the elastic neutron scattering experiments for \urs under the zero-stress-cooling condition, for the first time.

The single crystalline \urs was grown by using a tri-arc furnace, and vacuum-annealed at 1000$^{\circ}$C for a week. A plate-shaped sample with the dimensions of $\sim$ 25 mm$^2$ $\times$ 1 mm was cut from the crystal so that its base gives the (110) plane. It was put between the pistons (Be-Cu alloy) in the pressure cell, which is attached to the bottom of the insert of a $^4$He cryostat. We first cooled the sample down to 1.4 K under the stress-free condition, and then applied $\sigma$ along the [110] direction up to 0.4 GPa with keeping the sample at the same temperature. The magnitude of $\sigma$ was controlled with an oil-pressure device installed on the top of the insert, which was connected with the upper piston by a movable rod. The elastic neutron scattering experiments were performed by the triple-axis spectrometer GPTAS (4G) located at the Japan Atomic Energy Research Institute. The neutron momentum $k=2.660\ {\rm \AA^{-1}}$ was chosen by using the (002) reflection of Pyrolytic Graphite (PG) for the monochrometer and analyzer. We used the combination of 40'-80'-40'-80' collimators, together with two PG filters to eliminate the higher order reflections. The scans were performed in the $(hhl)$ scattering plane. The antiferromagnetic Bragg reflections were obtained by the longitudinal scans at the (111) position. We confirmed by the scans at 40 K for $\sigma=0$ that the contamination of the higher-order reflections can be neglected within the experimental accuracy. 

Figure 1 shows the (111) magnetic Bragg-peak intensities at 1.4 K for the increasing and decreasing $\sigma$ sweeps. The extrinsic angular-independent contributions on the intensities were subtracted from the data.  By increasing $\sigma$, the peak intensity is strongly enhanced. In addition, we found that the peak intensities for the $\sigma$-increasing process are significantly smaller than  those for the $\sigma$-decreasing process.
The difference can be more clearly seen by displaying the $\sigma$ variations of the AF moment \moment (Fig. 2 (a)). We here estimate the magnitudes of \moment by simply assuming the homogeneous AF phase, from the integrated intensities of the (111) magnetic Bragg peak normalized by the intensities of the weak nuclear (110) reflection. The $\mu_{\rm o}(\sigma)$ curve shows a clear hysteresis loop. \moment develops linearly from 0.016(4) \mb ($\sigma=0$) to 0.20(1) \mb ($\sigma=0.4\ {\rm GPa}$) with increasing $\sigma$. On the other hand, \moment for the $\sigma$-decreasing process shows nearly $\sigma$-independent behavior between 0.4 and 0.3 GPa, and then starts decreasing with further decompression. After the cycle of the compression, \moment returns to the initial value at ambient pressure. The $\mu_{\rm o}(\sigma)$ curve for the $\sigma$-decreasing process is quite similar to that obtained under the stress cooling condition~\cite{rf:Yoko2002}.

The observed magnetic Bragg-peak widths (FWHM) are slightly larger than the instrumental resolution, which are estimated from the widths of the higher-order reflections measured at the corresponding reciprocal lattice position without PG filters. We have estimated the correlation length of \moment along the [111] direction $\xi$, by fitting the observed (111) magnetic Bragg peaks with the Lorentzian function convoluted with the Gaussian resolution function. At ambient pressure, $\xi$ is estimated to be about 340 ${\rm \AA}$. The $\xi(\sigma)$ curve also shows a hysteresis loop (Fig.2 (b)), though detailed analyses are difficult because of the large experimental errors. The observed hysteresis loops in both $\mu_{\rm o}(\sigma)$ and $\xi(\sigma)$ curves strongly suggest that the $\sigma$-induced AF phase is in a metastable state, separated from the hidden ordered phase by a first-order phase transition.
\begin{figure}[tbp]
\begin{center}
\includegraphics[width=0.85\textwidth]{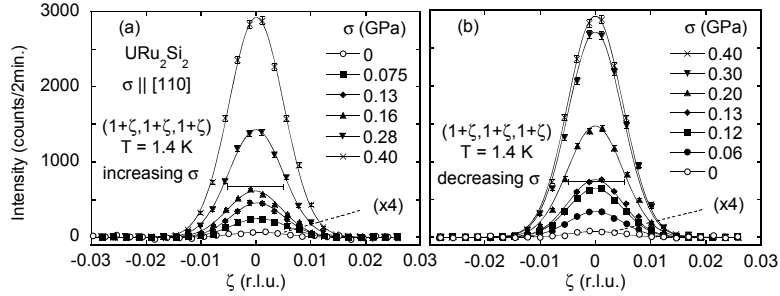}
\end{center}
\caption{The uniaxial stress variations of the magnetic Bragg peak intensities at 1.4 K for (a) $\sigma$-increasing process and (b) $\sigma$-decreasing process, obtained from the longitudinal scans around the (111) position. The horizontal bars indicate the widths of the resolution limit estimated from the higher-order nuclear reflections. Note that the data for $\sigma=0$ are 4 times enlarged.}
\end{figure}
\begin{figure}[tbp]
\begin{center}
\includegraphics[width=0.85\textwidth]{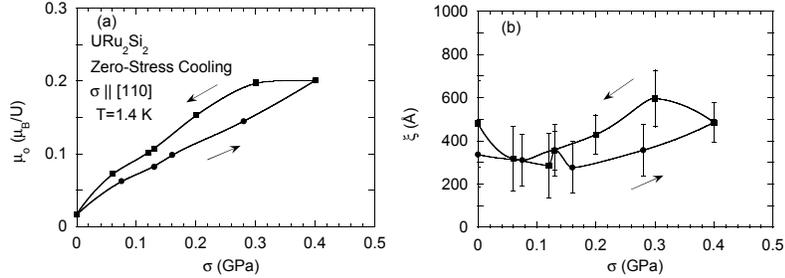}
\end{center}
\caption{The uniaxial stress variations of (a) the antiferromagnetic moment \moment and (b) the correlation length along the [111] direction $\xi$ for \urs , measured at 1.4 K after cooling the sample at $\sigma=0$. The solid lines are guide to the eyes.}
\end{figure}

The observed non-equilibrium AF state under the uniaxial stress implies that the AF moment is not directly coupled with the hidden order parameter. We now propose that the $c/a$ ratio ($\equiv \eta$) may govern the competition between these two order parameters. We previously found a clear anisotropy of the $\sigma$ dependence of the AF state: \moment is strongly enhanced for $\sigma$\,$\bot$\,[001] while nearly constant for \sigmatoc ~\cite{rf:Yoko2002}. The data give the information about the effects of $\sigma$ on the crystalline strains as follows. The uniaxial stresses are coupled with the strains by the elastic constants. By assuming the reported elastic constants for \urs ~\cite{rf:Wolf94}, we can thus estimate the increasing rates, $\partial \ln \eta /\partial \sigma$, to be $\sim 3.0 \times 10^{-3}$ GPa$^{-1}$ for both \sigmatoa and [110] within the linear approximation. Interestingly, the $\eta$ value calculated for hydrostatic pressure also increases with the rate $\partial \ln \eta /\partial P \sim 1.2\times 10^{-3}$ GPa$^{-1}$. The experimental results show the relationship, $\partial \mu_{\rm o}/\partial \sigma \sim 4 \times \partial \mu_{\rm o}/\partial P$~\cite{rf:Ami99,rf:Yoko2002}, which is in good agreement with the relationship between the calculated $\eta$ values, $\partial \ln \eta/\partial \sigma \sim 3 \times \partial \ln \eta /\partial P$. This means that both $\mu_{\rm o} (P)$ and $\mu_{\rm o}(\sigma)$ can be scaled by an implicit parameter $c/a$, at least, in the weak pressure range.

In summary, we have examined the effects of uniaxial stress on the AF state in \urs under the zero-stress-cooling condition for the first time, and observed the clear hysteretic behavior in the AF Bragg-peak intensities between the increasing and decreasing $\sigma$ sweeps. This indicates that the evolution of the AF phase induced by $\sigma$ is caused by the first-order phase transition. We also suggest from the analyses of the crystalline strain effects that the $c/a$ ratio is tightly coupled with the competition between the AF phase and the hidden ordered phase.

\end{document}